\documentclass{caosp306}

\usepackage{graphicx}

\usepackage{natbib}
\articleNo{000}
\pubyear{2018}
\volume{00}
\volnumber{0}
\firstpage{1}
\received{October 19, 2018}
\accepted{November xx, 2018}

\def\BibTeX{{\rm B\kern-.05em{\sc i\kern-.025em b}\kern-.08em
             T\kern-.1667em\lower.7ex\hbox{E}\kern-.125emX}}

\newcommand{\teff}{\ensuremath{T_{\rm eff}}}
\newcommand{\logg}{\ensuremath{\log g}}
\newcommand{\msun}{\ensuremath{\mathrm{M}_\odot}}
\begin{document}
\htitle{Elemental Abundance Analysis of Late-B Type Stars Using Sub-meter Class Telescopes}
%
\hauthor{T.\,K{\i}l{\i}\c{c}o\u{g}lu, R.\,Monier and E.\,Griffin}

\title{Elemental Abundance Analysis of Single and Binary Late-B Stars Using Sub-meter Class Telescopes: HR\,342, HR\,769, HR\,1284, and HR\,8705}


%
\author{
        T.\,K{\i}l{\i}\c{c}o\u{g}lu \inst{1} 
      \and 
        R.\,Monier \inst{2}   
       }

%
\institute{
           Ankara University, Faculty of Science, Department of Astronomy and Space Sciences, 06100,  Ankara, Turkey, \email{tkilicoglu@ankara.edu.tr}
         \and 
           LESIA, UMR 8109, Observatoire de Paris et Universit\'e Pierre et Marie Curie Sorbonne Universit\'es, place J. Janssen, Meudon, Paris, France 
          }
          
\date{October 19, 2018}

\maketitle

\begin{abstract}
We test the capabilities of 0.4 m telescopes equipped with an \'{e}chelle spectrograph to derive fundamental parameters and elemental abundances of four late-B type stars: HR\,342, HR\,769, HR\,1284, and HR\,8705. The medium resolution (R$\sim14\,000$) spectra covering the wavelength range of 4380-7350\,\AA~ of the four stars have been obtained using the 40-cm-telescope in Ankara University Kreiken Observatory (AUKR). Using spectrum synthesis, we were able to derive the abundances of eleven chemical elements. We find that these stars do not show remarkable departures from the solar abundances, except for HR 8705 and the primary component of HR 1284, which exhibit slight underabundances of a few elements, i.e., O, Mg, Al, Si, and Fe. We also find that HR 1284 is probably a new spectroscopic binary star. In order to model the spectrum of this object, one of us (TK) has developed a new graphic interface which allows us to synthesize the composite spectrum of binary stars.

\keywords{chemical abundance analysis -- chemically peculiar stars -- early type stars -- stars: individual: HR 342, HR 769, HR 1284, HR 8705}
\end{abstract}

%
\section{Introduction}
\label{intr}

The atmospheres of chemically peculiar (CP) late-B stars are useful laboratories to test the theory of atomic diffusion. We have recently started a project aiming at observing and analysing high resolution spectra of about 100 slowly rotating late-B stars in order to find new CP stars. In order to decide which objects should be monitored at a higher resolution with larger telescopes, we are currently  observing these objects using a 40-cm telescope equiped with a medium resolution \'{e}chelle spectrograph. We present here the abundance analysis of HR\,342, HR\,769, HR\,1284, and HR\,8705 and the detection of a new binary object (HR\,1284). We also present a new graphic interface which allows us to synthesize the spectrum of SB2 binaries.

\section{Observations and Analysis}
\label{obs}

\begin{figure}
\centerline{
\includegraphics[scale=1]{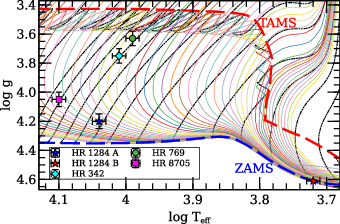}
\includegraphics[scale=1]{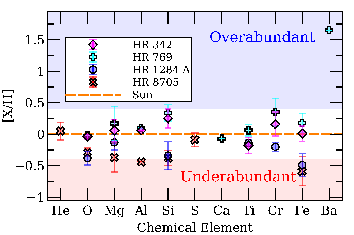}
}
\caption{Left: Stars on theoretical \logg{}--log\,\teff{} diagram (tracks taken from \citealt{bressanetal12}), Right: Abundance pattern of the stars.}
\label{abun}
\end{figure}

The targets have been observed using the Shelyak eShel spectrograph mounted on the 40-cm telescope in Ankara University Kreiken Observatory (AUKR) in 2017. The spectra span the spectral range 4380-7350\,\AA\ with a medium resolution, i.e., R$\sim14000$.
The atmospheric parameters were initially estimated from Johnson (BV) magnitudes with the calibrations of \citet{flower96} and then refined by modelling $H_\mathrm{\beta}$ lines.
The fundamental parameters of the stars were also estimated from \logg{}--\teff{} diagram (Fig. \ref{abun}, left) and are collected in Table \ref{funda}.

The model atmospheres were calculated using {\small ATLAS12} \citep{kurucz05} and the synthetic spectra were computed using {\small SYNSPEC49/SYNPLOT} \citep{hubenylanz95}. The linelist was first constructed from R. Kurucz's gfall.dat and then updated by using {\small VALD}, {\small NIST}, and recent publications.  We iteratively adjusted the synthetic spectra to the observed spectra until the best fit was achieved in order to derive the elemental abundances.

\begin{figure}
\centerline{\includegraphics[width=1.00\textwidth,clip=]{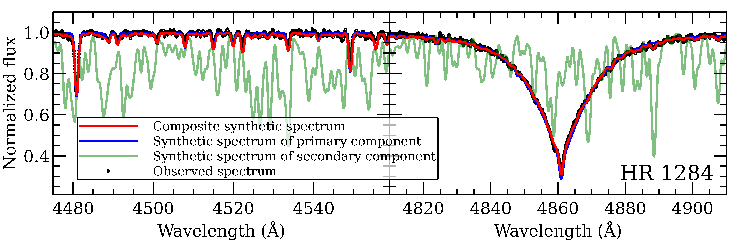}}
\caption{Observed spectra of HR\,1284 and synthesized spectra for its components.}
\label{sample}
\end{figure}

During the analysis, we have found that HR\,1284 most likely is a new SB2 star, where the lines of the secondary component are barely visible. One of us (TK) modified {\small SYNPLOT} into a new interface (called {\small SYNPLOTBIN}) to model the composite spectrum of this binary star. The observed spectrum is compared to the composite synthetic spectrum together with the synthetic spectra of each component in Fig. \ref{sample}. 

\section{Results and Conclusion}
\label{res}

We have derived the abundances of 11 elements for four bright stars with uncertainties ranging from $\pm0.10$ and $\pm0.25$\,dex
(Fig \ref{abun}, right). The chemical patterns of the stars do not depart significantly from the solar composition except for HR\,8705 and the primary component of HR\,1284, which exhibit slight underabundances of a few elements, such as O, Mg, Al, Si, and Fe. Observations of these two stars at a higher resolution on larger telescopes will help clarify their natures. This demonstrates the usefulness of small telescopes equipped with medium resolution spectrographs to derive abundances for a few (abundant) elements and sort out putative candidates for CP stars to be observed afterwards at higher resolution with meter class telescopes.

\begin{table}[t]
\small
\begin{center}

\caption{Fundamental parameters of the targets}
\label{funda}
\begin{tabular}{llrllr}
\hline\hline
Star  & Sp.T. & \teff{} (K) & \logg{} (cgs.) & $M$ (\msun) & Age (Myr) \\\hline
HR\,342     & B9.5V & $10250\pm250$  & $3.75\pm0.05$ & $3.00\pm0.10$ & $325\pm20$ \\
HR\,769     & B9.5V & $9800\pm200$  & $3.63\pm0.05$ & $3.05\pm0.10$ & $340\pm20$ \\
HR\,1284\,A & B9.5V & $11000\pm300$  & $4.20\pm0.05$  & $2.67\pm0.10$ & $200\pm20$ \\
HR\,1284\,B & K1V & $5200\pm300$  & $4.61$  & $0.90\pm0.02$ & $200\pm20$ \\
HR\,8705    & B8V & $12500\pm500$  & $4.05\pm0.05$  & $3.40\pm0.10$ & $158\pm20$ \\
\hline\hline
\end{tabular}
\end{center}
\end{table}
\bibliography{tkilicoglu}
\end{document}